\documentclass[12pt]{iopart}

\pdfoutput=1

\usepackage{color}
\usepackage{cite}
\usepackage{url,psfrag,graphicx}
\usepackage{dcolumn}
\usepackage{bm}
\usepackage{hyperref}
\usepackage{float,epsfig,color}
\usepackage{times}
\usepackage{array}
\usepackage{braket}
\usepackage{diagbox}
\usepackage{lscape}
\usepackage{multirow}
\usepackage{stackrel}
\usepackage{verbatim}
\usepackage{xcolor}
\usepackage{soul}
\usepackage{hyperref}
\usepackage{iopams}
\usepackage{setstack}
\usepackage{orcidlink}

\usepackage{dcolumn}
\newcolumntype{.}{D{.}{.}{-1}}
\newcolumntype{(}{D{(}{(}{-1}}

\newcommand{\tavg}[1]{\langle {#1} \rangle}
\newcommand{\davg}[1]{\overline {\tavg{#1}} }
\newcommand{\SDY}{\mathcal{S}}
\newcommand{\ql}{\ensuremath{q_\mathrm{l}}}

\newcommand{\eqref}[1]{(\ref{#1})}
\DeclareMathAlphabet{\mathitbf}{OML}{cmm}{b}{it}

\begin{document}

\title[Cluster moves with an entropic reservoir accelerate low-temperature simulations]{Cluster moves with an entropic reservoir accelerate low-temperature simulations \\of three-dimensional spin glasses}

\author{C. Chilin$^{1,2}$\orcidlink{0009-0005-6012-4915}, E. Marinari$^{2,3}$\orcidlink{0000-0002-3464-4133}, V. Martin-Mayor$^{1}$\orcidlink{0000-0002-3376-0327}, G. Parisi$^{4,2,3}$\orcidlink{0000-0001-6500-5222}, J.J.~Ruiz-Lorenzo$^{5,6}$\orcidlink{0000-0003-0551-9891}, and D.~Yllanes$^{7,8,9}$\orcidlink{0000-0001-7276-2942}}

\address{$^1$ Departamento de Física Teórica, Universidad Complutense, 28040 Madrid, Spain}
\address{$^2$ Dipartimento di Fisica, Sapienza Università di Roma, 00185 Rome, Italy}
\address{$^3$ CNR-Nanotec, Rome Unit, and INFN, Sezione di Roma, 00185 Rome, Italy}
\address{$^4$ International Research Center of Complexity Sciences, Hangzhou International Innovation Institute, Beihang University, Hangzhou 311115, China}
\address{$^5$ Departamento de Física, Universidad de Extremadura, 06006 Badajoz, Spain}
\address{$^6$ Instituto de Computación Científica Avanzada (ICCAEx), Universidad de Extremadura, 06006 Badajoz, Spain}
\address{$^7$ Fundación ARAID, Diputación General de Aragón, 50018 Zaragoza, Spain}
\address{$^8$ Instituto de Biocomputación y Física de Sistemas Complejos (BIFI), Universidad de Zaragoza, 50018 Zaragoza, Spain}
\address{$^9$ Zaragoza Scientific Center for Advanced Modeling (ZCAM), 50018 Zaragoza, Spain}
\ead{clachili@ucm.es}

\begin{abstract}
We present an algorithm for the simulation of three-dimensional spin
glasses deep in the low-temperature phase: Parallel
Tempering enhanced with Houdayer moves and with an entropic reservoir
(PTHR). Although differences with the standard Houdayer algorithm
are small, PTHR allows us to equilibrate a large number of
samples of $L=16$ lattices with Gaussian couplings for temperatures
$T\geq 0.2$. We show that the computational complexity displays better
size scaling than standard Parallel Tempering. For finite sizes, our
method  outperforms other cluster algorithms by a speedup factor of around
64.  In close analogy with standard Parallel Tempering, PTHR's computational complexity  strongly relates to temperature chaos.
\end{abstract}


\section{Introduction}\label{sec:intro}

Spin glasses~\cite{mydosh:93,young:98} are magnetic disordered alloys that have aroused the curiosity of physicists from many points of view~\cite{mezard:87,charbonneau:23,parisi:23,dahlberg:25}.
In particular, connections with computer science have been emphasized, because finding the ground state of a spin-glass Hamiltonian in three spatial dimensions ---specifically, the Edwards-Anderson model~\cite{edwards:75,edwards:76} with Ising spins--- is an NP-complete problem~\cite{barahona:82b,istrail:00}.  Hence, simulating spin glasses at zero temperature, $T=0$, is a paradigmatic example of a  hard computational problem. Yet, equilibrating three-dimensional instances of the Edwards-Anderson model down to low temperatures is  even more difficult ---it is such a challenging task that
dedicated computers have been specifically built to address it~\cite{ogielski:85,cruz:01,janus:08,matsubara:20,mcgeoch:22,mcmahon:16,janus:12b,janus:14,sajeeb:26,zhu:26}. Although we lack  mathematical proof of the statement that sampling from the Boltzmann weight is more difficult at temperature $T>0$ than at $T=0$, there is empirical evidence for it that we review next.

A review of modern approaches for sampling at $T=0$ ---\emph{i.e.}, finding the ground state--- can be found in Ref.~\cite{caracciolo:22}. Most relevant for us are the 
sizes reached in large-scale simulations that found ground states for a significant number of problem instances ---also named samples. A sample is characterized by the set of couplings $\{J_{ij}\}$, where 
$J_{ij}$ is the strength of the pairwise interaction between spins $\sigma_i$ and $\sigma_j$; see Eq.~\eqref{eq:hamiltonian} below. In $T=0$ simulations, couplings are continuously distributed random variables (quite often the $J_{ij}$ are Gaussian-distributed random variables, with zero mean and variance $J$; one typically chooses energy units such that $J=1$). Exact-solver simulations include Ref.~\cite{palassini:00}, in lattices
with linear sizes $L\leq 8$; Ref.~\cite{palassini:03}, which reached $L=10$ (although with open boundary conditions); and Ref.~\cite{shen:24}, which reached $L=12$ 
with periodic boundary conditions. On the other hand, we have heuristic-solver simulations, \emph{i.e.}, simulations where the ground state was reached with high 
probability.  In this category, we find Ref.~\cite{pal:96}, which reached $L\leq 10$, while Ref.~\cite{roma:09} went up to $L\leq 11$. The absolute record to date is  in Ref.~\cite{marinari:01}, which obtained large statistics using a genetic algorithm in lattices $L\leq 14$. More recent work reaches good statistics for $L\leq 12$ and just some ten samples of $L=14$~\cite{delbono:26}, so we think it is fair to say that the absolute record still belongs to Ref.~\cite{marinari:01}. 

As for $T>0$ simulations, the two main algorithms currently on the market are Parallel Tempering (PT)~\cite{hukushima:96, marinari:98b} and Population Annealing (PA)~\cite{hukushima:03,machta:10,wang:15b}, of which the latter is more suited for massively parallel computations on a distributed architecture~\cite{barash:17}.
Both algorithms are very sensitive to the minimal temperature reached in the simulation, $T_\mathrm{min}$, in the sense that the lower $T_\mathrm{min}$ the smaller 
the size $L$ at which one manages to equilibrate. Consider, for instance, the PT simulations of Refs.~\cite{janus:10,janus:10b} with binary couplings (\emph{i.e.}, $J_{ij}=\pm 1$ with 50\% probability). These authors went down to $T_\mathrm{min}=0.125\approx 0.11 T_\mathrm{c}$  on lattice sizes $L\leq 8$, where $T_\mathrm{c}$=1.102(3) is the critical temperature~\cite{janus:13}.\footnote{It was felt that the 
binary couplings significantly distorted the physics at such a low $T_\mathrm{min}$.} For $L=12$, however, they had to stop at $T_\mathrm{min}=0.414\approx 0.38 
T_\mathrm{c}$ and for larger sizes, using the Janus special-purpose computer~\cite{janus:12b}, they reached $T_\mathrm{min}=0.479\approx 0.44 T_\mathrm{c}$ ($L=16$), 
$T_\mathrm{min}=0.625\approx 0.57 T_\mathrm{c}$ ($L=24$), and
$T_\mathrm{min}=0.703\approx 0.64 T_\mathrm{c}$ ($L=32$). At very low temperatures, Gaussian-distributed couplings, which result in $T_\mathrm{c}=0.95(4)$~\cite{marinari:98d}, are usually regarded as preferable. For this version of the model
we have two PT simulations that went down to $T_\mathrm{min}=0.2\approx 0.21 T_\mathrm{c}$ ($L=8$~\cite{katzgraber:01} and 
$L=10$~\cite{katzgraber:07}). On the PA side, we have $T_\mathrm{min}=0.01\approx 0.011T_\mathrm{c}$ ($L=10$~\cite{wang:26}),
$T_\mathrm{min}=0.333\approx 0.35 T_\mathrm{c}$ ($L=12$~\cite{wang:17}), and $T_\mathrm{min}=0.42\approx 0.44 T_\mathrm{c}$ ($L=16$~\cite{wang:20}). 
At least in the case of PT~\cite{janus:10,fernandez:13,billoire:18} ---but we strongly suspect this is also the case with PA---  
the physical mechanism of temperature chaos, see Sec.~\ref{sec:pthr-vs-trw}, underlies the decline of the algorithmic performance as $L$ grows.

The comparison of the above two paragraphs makes it obvious that the ground-state approach (\emph{i.e.},~$T_\mathrm{min}=0$) is leading the race towards large $L$, so 
far. This state of affairs is, however, unsatisfactory from the physics point of view, where one needs to take  the limit $L\to\infty$  \emph{first} and only 
afterwards consider what happens in the $T\to 0$ limit. This is quite crudely exemplified, for instance, in the analysis of two-dimensional spin glasses; see, \emph{e.g.},~\cite{fernandez:16b}. Therefore, we describe here how we have modified Houdayer's cluster 
method~\cite{houdayer:01}, see also Ref.~\cite{zhu:15}, to obtain a new algorithm that has allowed us to equilibrate a three-dimensional spin glass with 
Gaussian couplings down to  $T_\mathrm{min}=0.2\approx 0.21 T_\mathrm{c}$ on lattices $L\leq 16$. Also, we think that equilibration of $L =18$ is within reach 
algorithmically (it was barely out of our computational budget). Our approach to the discussion will be that of a working physicist: we need to simulate a 
large number of samples (around $2000$), down to low temperatures, with a computational budget of about $2.5\times 10^6$ CPU hours. Since we wanted to make the 
most out of our computing power, we employed multispin coding~\cite{jacobs:81} adapted to our Gaussian couplings, see~\ref{sec:appendice_tecnico}.
If one has significant GPU computing resources instead, some of the algorithmic decisions we made would perhaps not be optimal. Nevertheless, we hope that the detailed explanations 
below will help the reader optimize the algorithm also in that setting. 

Before entering into details, let us recall that developing efficient mappings 
of spin-glass simulations into graphical representations ---\emph{i.e.}, finding clusters of
spins to flip simultaneously--- has been a holy grail for a long time~\cite{swendsen:86}. 
This interest in cluster moves is motivated by their great success in
ferromagnetic systems~\cite{wolff:89, swendsen:87}. When disorder comes into play, it is impossible to identify
the clusters just by looking at the spins of a single configuration, and one has to rely on
two-replica graphical representations, such as the Houdayer algorithm~\cite{houdayer:01}. 
Even though the ``right'' way to compute the cluster is
far from straightforward ---see, \emph{e.g.}, Refs.~\cite{machta:08, munster:26}---, cluster moves have proven useful
for models where frustration is present~\cite{keiser:26}. It is also interesting that two-replica graphical
representations have been shown to leave a sign at the spin-glass transition in three dimensions,
also when considering Houdayer clusters; see Refs.~\cite{machta:08} and~\cite{munster:26}.

The remainder of this work is organized as follows. We frame the problem and our approach in the extended 
introduction in Sec.~\ref{subsec:extended-Intro}. This section assumes from the reader some familiarity with the notion of spin overlap (local and global). Readers
needing a more paused exposition may prefer to read Sec.~\ref{sec:model_obs} first, which describes the Hamiltonian and the observables that we shall consider. Sec.~\ref{sec:true-algorithm} describes in detail the final form of our algorithm. Next, in Sec.~\ref{sect:comparison-algorithms}, the new 
algorithm is quantitatively compared with regular PT and with the version of the cluster algorithm from Ref.~\cite{zhu:15}. Sec.~\ref{sec:pthr-vs-trw}
explains that, in contrast with plain PT, hard-to-equilibrate samples cannot be identified from their temperature-diffusion dynamics. However, these hard-to-equilibrate samples ---hard even with the cluster method--- still present a particularly strong temperature chaos~\cite{billoire:18}.
Finally, Sec.~\ref{subsec:rtrips}, shows that we are still able to recognize a typical timescale from the PT temperature random walk that is characteristic of the equilibration of the whole set of samples. We present our conclusions in Sec.~\ref{sect:conclusions}. The paper is complemented 
with an \ref{sec:appendice_tecnico}  that contains the crucial parameters that characterize our simulations and some information about our multispin coding program.

\subsection{Computational challenge and simulation strategy}\label{subsec:extended-Intro} 

In this project we aim to equilibrate a large number of samples ($\approx 2000)$ with Gaussian-distributed couplings, down to temperature $T_\mathrm{min}=0.2\approx 0.21 T_\mathrm{c}$, low enough to allow for a smooth extrapolation to $T=0$. Of course, the larger the $L$ we reach, the better. 

Our first attempts in this direction used
the standard PT algorithm on more modern, faster CPUs than the authors of Refs.~\cite{katzgraber:01,katzgraber:07} had access to. These standard PT simulations of ours  passed our equilibration test 
of choice, described in Sec.~\ref{sec:model_obs},
down to $T_\mathrm{min}=0.2$ for systems of sizes $L\leq 12$. 
Actually, we were also able to equilibrate a small number of systems with $L=14$. It 
was clear at this point, however, that some algorithmic improvement was necessary if we were to expand the range of system sizes in our study~\cite{chilin:26}. 

An obvious candidate for an improved PT was adding Houdayer cluster updates to the standard single-spin-flip Metropolis~\cite{houdayer:01}. These cluster moves require 
simulating two replicas, since the updates modify both configurations simultaneously. The method was
originally proposed in 2D~\cite{houdayer:01}, where there is only a paramagnetic phase as soon as $T>0$ and clusters do not percolate. Indeed, the cluster move affords
a very significant  speedup in the thermalization process on the square lattice~\cite{fernandez:16b}. Unfortunately, it was clear from the 
outset~\cite{houdayer:01} that Houdayer clusters in 3D would percolate even at very high temperatures. Conventional wisdom had it that the lattice would simply split into two clusters. 

It is well known that, when clusters 
are too big, the cluster move becomes a simple symmetry transformation of the whole system, producing no algorithmic gain~\cite{sokal:97}.
In the case of spin glasses, when the Houdayer clusters are too big the two replicas simply exchange their 
configurations. 
This probably explains why the Houdayer method was 
largely ignored for simulations in space dimension $D>2$, although Ref.~\cite{zhu:15} suggested that a significant speedup can also be obtained in $D=3$.
We were initially skeptical, however, because the authors of~\cite{zhu:15} were able to equilibrate only systems with $L\leq 12$ down to temperatures
$T\geq T_{\mathrm{min}}=0.42$ which was not an improvement over our own simulations with regular PT~\cite{janus:10,billoire:18}. Nevertheless, 
having at our disposal a set
of configurations equilibrated with standard PT, we decided to study the geometry of the Houdayer clusters to decide whether an algorithmic gain was achievable.

\begin{figure}[t]
  \centering
  \begin{minipage}{0.49\linewidth}
    \centering
    \includegraphics[width=\linewidth]{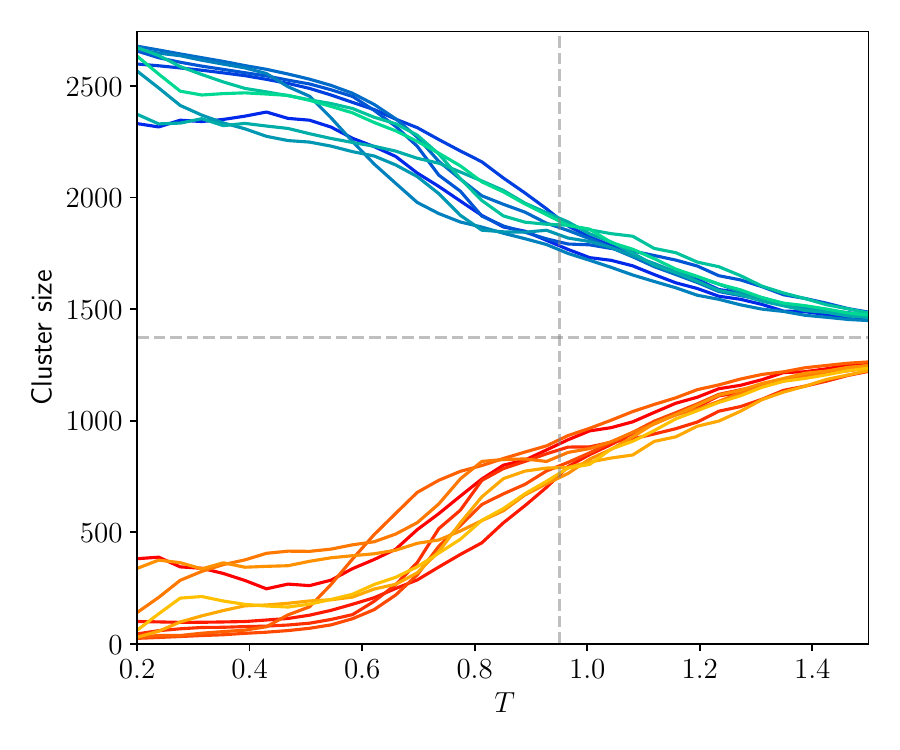}
  \end{minipage}
  \begin{minipage}{0.49\linewidth}
    \centering
    \includegraphics[width=\linewidth]{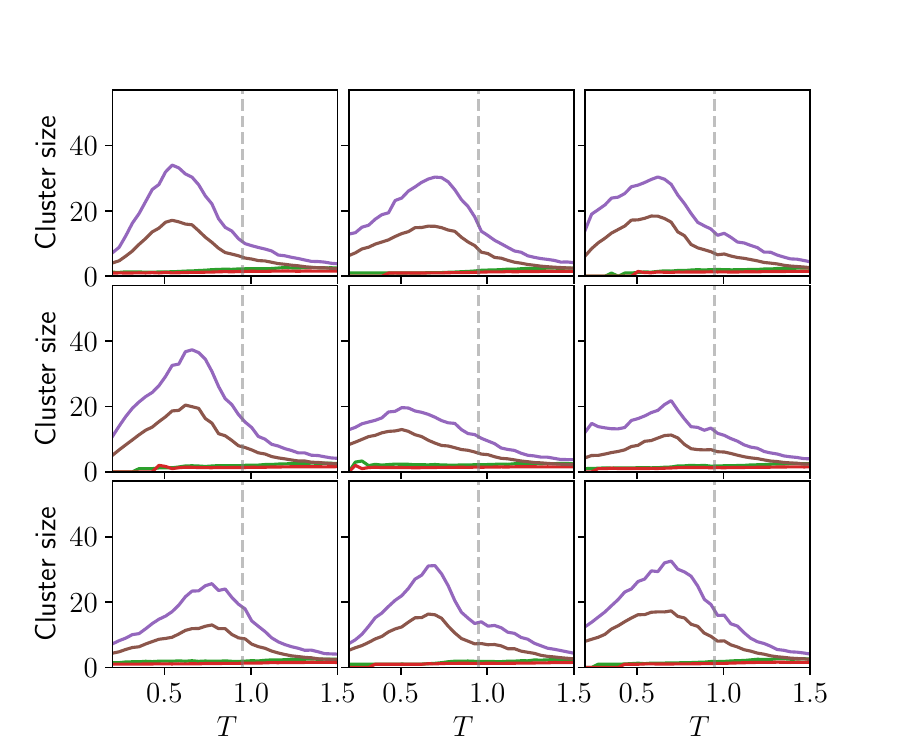}
  \end{minipage}

  \caption{\label{fig:cluster-zoology}
    \textbf{Largest clusters in equilibrated configurations.}
    Temperature dependence of the size of the six largest clusters that are present in a given pair of spin configurations, as computed for 9 different \(L=14\) samples. For each temperature, we present an average over equilibrated configurations. The vertical dashed lines correspond to \(T=0.95 \approx T_\mathrm{c}\)~\cite{marinari:98d}, 
    and the horizontal ones indicate half the size of the system \(V/2=1372\).
    \textbf{Left:} The blue-shaded lines in the upper half correspond to the biggest clusters of the majority (\emph{i.e.}, clusters with the same overlap sign as the full lattice). There are nine curves (one per sample).
    The orange-shaded curves are for the biggest cluster of the minority.
    \textbf{Right:} Average size of the 
    second-largest cluster of the majority (purple) and the minority (brown) for each of the 9 samples. As a comparison, also shown are the average sizes of the third cluster of the majority (green) and the minority (red).
  }
\end{figure}

Let us recall that Houdayer's clusters divide the lattice into connected components: two neighboring lattice sites $i$ and $j$ belong to the same connected component if  (and only if) $q_i=q_j$ [$q_i$ and $q_j$ are the respective site overlaps, see Eq.~\eqref{eq:site_ovlp}]. Hence, if the global overlap defined in Eq.~\eqref{eq:tot_ovlp} is $q>0$ ($q<0$), the majority of sites will belong to 
clusters with $q_i=1$ ($q_i=-1$). Let us use  \emph{majority} to refer to the clusters with $q_i  q>0$ (the \emph{minority} will refer to the clusters with $q_i  q<0$), that is,  the majority clusters have the same overlap sign as the full lattice.

Now, see 
Fig.~\ref{fig:cluster-zoology}, we found that, at low temperatures, the largest cluster in both the majority and the minority contained a significant fraction of the lattice. Hence,
flipping the largest cluster, either from the majority or from the minority, essentially amounts to a useless symmetry transformation.
Interestingly enough, however, the second-largest cluster in both sets contains a few dozen lattice sites whose flipping was potentially useful. This is 
why we decided to opt for an all-cluster version of the Houdayer algorithm, analogous to the Swendsen-Wang form~\cite{swendsen:87} of the cluster algorithm 
for ferromagnetic systems; Refs.~\cite{houdayer:01,zhu:15} followed instead the analogue of the single-cluster approach~\cite{wolff:89}.

A completely unexpected ---and somewhat undesirable--- consequence of the cluster move, see Sec.~\ref{sec:pthr-vs-trw} below, was that our favourite equilibration 
test~\cite{fernandez:09b,janus:10,billoire:18} becomes useless (the test focuses on the temperature diffusion of the different clones in the PT). This 
motivated an additional modification of the algorithm, which we name the \emph{entropic reservoir} and is explained in Sec.~\ref{sec:true-algorithm}. Thanks to this enhanced version of PT we have equilibrated systems down to $T_\mathrm{min}=0.2$ up to the size of $L=16$ and nearly equilibrated $L=18$ samples, see \ref{sec:appendice_tecnico}. 

For the sake of completeness, let us mention that Houdayer's cluster move generates a very significant speed up in a $D=3$ lattice with the geometry of an 
elongated right prism. Indeed, at low $T$ the clusters split the prism into several very large but not percolating connected components~\cite{bernaschi:26b}.

\section{Model and observables}\label{sec:model_obs}
We study the three-dimensional Edwards-Anderson spin glass, which is defined by the Hamiltonian~\cite{edwards:75, edwards:76}
\begin{equation} \label{eq:hamiltonian}    
\mathcal{H} \left\{ {\mathbf{\sigma}} \right\} = -\sum_{ \langle i, j \rangle } J_{ij}\, \sigma_i\, \sigma_j\,,
\end{equation}
where the sum runs over all pairs of nearest neighbours $ \langle i, j \rangle $ in a 3D simple-cubic lattice of linear size $L$ with periodic boundary conditions. We denote the total number of sites in the lattice by $V=L^3$. The Ising-like variables $\sigma_i=\pm 1$ that form the configuration $\mathbf{\sigma}= \{ \sigma_1,\dots , \sigma_V \}$ are termed \textit{spins}. In our case, the couplings $J_{ij}$ are randomly extracted from a Gaussian distribution with zero mean and unit variance. For this version of the model, the spin-glass transition takes place 
at $T_\mathrm{c}=0.95(4)$~\cite{marinari:98d}. During the study of each \textit{sample} ---a specific instance of the couplings--- the couplings are fixed once for all at the beginning of the simulation (quenched disorder).  We first perform the thermal average of all interesting  observables $\mathcal{O}$, denoted as $\tavg{\mathcal{O}}$, and afterwards we carry out the average over the disorder ---\emph{i.e.}, over the different samples---, which is referred to as $\davg{ \mathcal{O}}$.
The Hamiltonian in Eq.~\eqref{eq:hamiltonian} enjoys a global $Z_2$ symmetry: in the absence of an external magnetic field, the energy does not change if all spins are flipped, \emph{i.e.},  $\sigma_i \longrightarrow -\sigma_i$ for all $i$.

As usual in spin glasses, the important observable quantities are defined in terms of real replicas: copies of the system that evolve under the same couplings $\{J_{ij}\}$ but are otherwise statistically independent. We define the \textit{site spin overlap} of site $i$ between spins belonging to replicas $(a)$ and $(b)$ as 
\begin{equation} \label{eq:site_ovlp}
    q_i = \sigma_i^{(a)} \sigma_i^{(b)}\,.
\end{equation}
The global \textit{spin overlap} ---or sometimes simply \textit{overlap}--- is given by the average of the site overlap over sites:
\begin{equation}\label{eq:tot_ovlp}
    q=\frac{1}{V} \sum_{i=1}^{V} q_i\,,
\end{equation}
and is regarded as the order parameter for the spin-glass phase transition. Also of interest will be the \textit{link overlap} 
\begin{equation}\label{eq:link_ovlp}
    q_\mathrm{l} = \frac{1}{3  V} \sum_{\langle i, j \rangle} \sigma_i^{(a)} \sigma_i^{(b)} \sigma_j^{(a)} \sigma_j^{(b)} \,.
\end{equation}
The physical interest of the link overlap is discussed, for instance, in Refs.~\cite{krzakala:00,palassini:00,janus:10}. Here, we need to consider it because it is crucial for our equilibration test, which exploits a relation of Schwinger-Dyson type~\cite{bray:80c}. Since the identity holds \emph{only} in thermal equilibrium, its fulfilment serves as a very sensitive equilibration 
test~\cite{katzgraber:01}. Specifically, the Schwinger-Dyson-Young (SDY) observable is
\begin{equation}\label{eq:SDY-def}
   {\cal S} =  \frac{\cal H}{3 V} + \frac{(1-q_\mathrm{l})}{T}\,,
\end{equation}
where we have specialized to Ising spins on a simple-cubic lattice with Gaussian-distributed couplings. The Schwinger-Dyson-like identity that we are referring to is
\begin{equation}\label{eq:SDY}
   \overline{\langle {\cal S}\rangle} =0 \,. 
\end{equation}
A well-equilibrated sample may have ---and indeed generally has--- $\langle S\rangle\neq 0$. Only after the disorder average does one get zero.  Alternatively, a positive value of $\davg{\SDY}$ is a telltale sign of lack of equilibration, because the expected values of both ${\cal {H}}$ and $(1-\ql)$ decrease as the thermalization process proceeds.

\section{Description of the algorithm}\label{sec:true-algorithm}

Our procedure resembles the one proposed by Houdayer~\cite{houdayer:01} for the 2D Ising spin glass, but with a modification to the cluster move and 
the addition of an \textit{entropic reservoir}  on top of PT. 

For the PT procedure we run $N_T$ simulations at uniformly spaced temperatures $T_k$. In order to perform a cluster move, we need at least $N_\mathrm{r}=2$ 
replicas of the system for each temperature. We can see this as having $N_\mathrm{r}$ groups of $N_T$ configurations. We refer to the configurations in the same ``replica group'' as \textit{clones}, while we will refer to configurations belonging to different replica groups as \textit{replicas}. Unless otherwise specified, when we 
refer to replicas we mean configurations belonging to different replica groups but at the same temperature.

An elementary Monte Carlo step (EMCS) develops as follows:
\begin{enumerate}
    \item 10 full-system updates with local moves at fixed temperature;
    \item 1 all-cluster update (ACU) step;
    \item 1 temperature exchange (or PT) step;
    \item 1 entropy-reservoir (ER) step.
\end{enumerate}

\emph{The full-system update at fixed temperature} is performed independently for every replica and for every clone. Every spin in the lattice is given a chance to be flipped during the full-system sweep. In this work, a sweep consists in the  sequential update of the whole lattice in lexicographic order.  
A common choice is to use Metropolis updates for the spins, but other algorithms such as a heat bath could be used. Most of our simulations were conducted with Metropolis but we also employed  Microcanonical Simulated
Annealing (MicSA)~\cite{bernaschi:26}, which is slower in terms of convergence towards the equilibrium 
distribution but is faster in terms of CPU time.
The choice of 10 full-system updates comes from a balance between the individual computational 
cost of the steps of the algorithm.

\emph{The all-cluster update} (ACU) involves the two clones from replicas  $a$ and $b$ that occupy a given temperature $T_k$ in the PT grid. One can use any number of replicas $N_\mathrm{r}$ and divide them in pairs $(a,b)$, but in this work we stick to $N_\mathrm{r}=2$.
The move consists of the following steps: First, we compute the site overlap [Eq.~\eqref{eq:site_ovlp}] for each site $i$ of the system. Second, the lattice
is divided into connected components (or clusters). Two lattice nearest neighbours $i$ and $j$ belong to the same cluster if, and only if, $q_iq_j>0$.
Third, we independently choose a random sign $\eta_c=\pm 1$ for every cluster, with 50\% probability. All the spins belonging to cluster $c$, in both replicas, are multiplied by $\eta_c$ (if site $i$ belongs to cluster $c$, then $\sigma_{i,a}\rightarrow \eta_c \sigma_{i,a}$ and $\sigma_{i,b}\rightarrow \eta_c \sigma_{i,b}$).

Let us now argue that the cluster move is symmetric in probability with its reversed move. The crucial point is that the local overlaps are not 
modified by the cluster move,  
$q_i \rightarrow \eta_c \sigma_{i,a}\,\eta_c \sigma_{i,b}=q_i$.  Now,  the $q_i$ being 
identical, the lattice connected 
components are the same for the original and transformed configurations. Thus, the probability of choosing a set of signs  $\{\eta_c\}$ is identical 
for the direct and for the reversed cluster move. And,
if one chooses again the same set of signs $\{\eta_c\}$ for the final configuration, the original configuration is recovered.

Furthermore, the cluster move leaves the total Hamiltonian  
$\mathcal{H}=\mathcal{H}^{(a)}+\mathcal{H}^{(b)}$ (\emph{i.e.}, the Boltzmann weight of the replica pair) unchanged. Therefore,  being  symmetric in probability, the ACU step of our algorithm verifies detailed balance; see, for instance,  Ref.~\cite{sokal:97}.

In order to save computer time,  we restrict the ACU move to clones occupying temperatures $T<1$. This choice relates to the temperature dependence of the size of the second- and third-largest clusters. As shown in Fig.~\ref{fig:cluster-zoology}, these clusters reach a maximum size for a sample-dependent temperature and then rapidly shrink as $T$ increases above $T_\mathrm{c}$. Our choice $T<1$ includes the maximum of the second-largest cluster in all the cases that we have examined.

Let us remark the two key differences with Houdayer's original proposal~\cite{houdayer:01}. First, we flip clusters with both signs of the overlap while in the 
original algorithm only clusters with negative overlap were considered. Of course, Houdayer was aware that positive-overlap clusters can be 
flipped as well~\cite{houdayer:01}, but he wanted to keep the algorithm valid also when a magnetic field is applied (in a field, the cluster move keeps the 
total energy unchanged only if the overlap is negative). Second, in the original algorithm only one cluster is traced and then flipped with probability one, in close analogy with Wolff's single-cluster method~\cite{wolff:89}. We think ACU is preferable in our case because whatever the speedup that the cluster move will afford, it will come from the second- and third-largest clusters, which would rarely be picked in a single-cluster update.

\emph{The PT step} is conducted independently for each replica. Each clone in a replica occupies a different temperature in the PT grid, which  is described in \ref{sec:appendice_tecnico}. In the PT step, the clones that occupy  neighbouring temperatures  in the grid, 
$T_k<T_{k+1}$,  attempt to exchange their temperature, a move that is accepted with a rule that verifies detailed 
balance~\cite{hukushima:96,marinari:98b,marinari:98g}. 

The rationale of the PT move is that every clone alternates time lapses of quick decorrelation ---when the clone
occupies the highest temperatures in the grid--- with periods of exploration of the nearest free-energy minimum ---when the clone occupies low temperatures. For later reference, let us remark that every clone performs a random walk over the temperature grid, which we shall refer to as the $T$ random walk (TRW). In a sane PT simulation the TRW should be ergodic over the temperature grid.

\emph{The ER step} is carried out independently for each replica. This step involves the use of additional clones at $T=T_\mathrm{max}$.
When a clone in the PT hits the highest temperature $T_\mathrm{max}$, it enters the ER, where it runs for a number of Metropolis steps high enough to get an independent spin configuration. After its spin configuration has been refreshed at $T_\mathrm{max}$, the clone is exchanged with the clone that occupies the highest
temperature $T=T_\mathrm{max}$ in the PT. In this way, the clone  that has just exited the ER could have its temperature lowered during the PT step of the algorithm. Should the clone stay at $T_\mathrm{max}$, it would be pushed back into the ER to be refreshed again. 

In order to determine how long a clone should stay in the ER before it is reinjected into the PT, we have computed the time-dependent spin 
correlation function 
\begin{equation}
C(\tau) = \frac{1}{V} \sum_i \davg{\sigma_i(t) \sigma_i(t+\tau)}\,,
\end{equation}
where the starting configuration $\{\sigma_i(t)\}$ is in thermal equilibrium at temperature  $T_\mathrm{max}$ and the final configuration
$\{\sigma_i(t+\tau)\}$ results from $\tau$ full-lattice sweeps at fixed temperature $T_\mathrm{max}$. The two limiting values are $C(\tau=0)=1$, \emph{i.e.}, perfect memory of the starting configuration, and $C(\tau\to\infty)=0$, which indicates complete memory loss. We have obtained $C(\tau=960)< 10^{-3}$ irrespective of the system size. Hence, we have fixed 960 steps as the standard time that a clone should spend in the ER.

For technical reasons, related to our multispin coding implementation, we kept 4 clones per replica within the ER. While the clones in the PT perform an EMCS, the clones in the reservoir are evolved for $960/4$ Metropolis steps in a single reservoir update. The clone that gets injected back into the PT is the one that has spent the longest time within the reservoir.

The rationale for the ER is that, as we shall show below and in sharp contrast with plain PT 
simulations~\cite{fernandez:09b,janus:10,billoire:18}, the mixing time for the temperature random walk is no longer the characteristic timescale for 
reaching thermal equilibrium. Rather, a single clone makes many round-trip excursions from $T_\mathrm{max}$ to $T_\mathrm{min}$ and back to $T_\mathrm{max}$ 
before the equilibration criterion in Eq.~\eqref{eq:SDY} is met. Therefore, we consider it important to ensure that whatever clone reaches $T_\mathrm{max}$ refreshes its  spin configuration. Our choice of parameters ensures that the new configuration is statistically independent of the previous one and is well equilibrated  at $T_\mathrm{max}$.

\section{Comparison with other algorithms}\label{sect:comparison-algorithms}
The algorithm proposed here will be referred to as PTHR, an acronym for PT enhanced with Houdayer moves and with an entropic Reservoir. We shall compare PTHR with two other algorithms. We shall start by confronting our algorithm with the standard PT at $L=14$ down to  $T_\mathrm{min}=0.2$. Next, we shall compare PTHR with the cluster-amplified PT algorithm of Ref.~\cite{zhu:15}, which we shall name PT-ICM following these authors. The comparison between PTHR and PT-ICM will be carried out under the conditions chosen for the benchmark in 
Ref.~\cite{zhu:15}, namely $L=12$  and $T_\mathrm{min}=0.42$.

Since reaching thermal equilibrium in spin glasses is a demanding task, it is customary to study the equilibration process by considering a logarithmic timescale. Let $t_\mathrm{max}$ be the maximum number of EMCSs carried out in the simulation. We split the Monte Carlo time series by averaging observables over logarithmic time bins ($\log_2$-binning)
\begin{equation}\label{eq:time-bin-def}
\mathrm{bin}(b)=\Big[\frac{t_\mathrm{max}}{2^{b+1}}+1, \frac{t_\mathrm{max}}{2^{b}}\Big]\,.
\end{equation}
Therefore, bin  0 coincides with the second half of the simulation, bin  1 is the second quarter, bin  2 is the second eighth, and so on. It will be useful to compare the average of the different observables as computed for the  different time bins.

\begin{figure}[h]
    \centering
    \includegraphics[width=\linewidth]{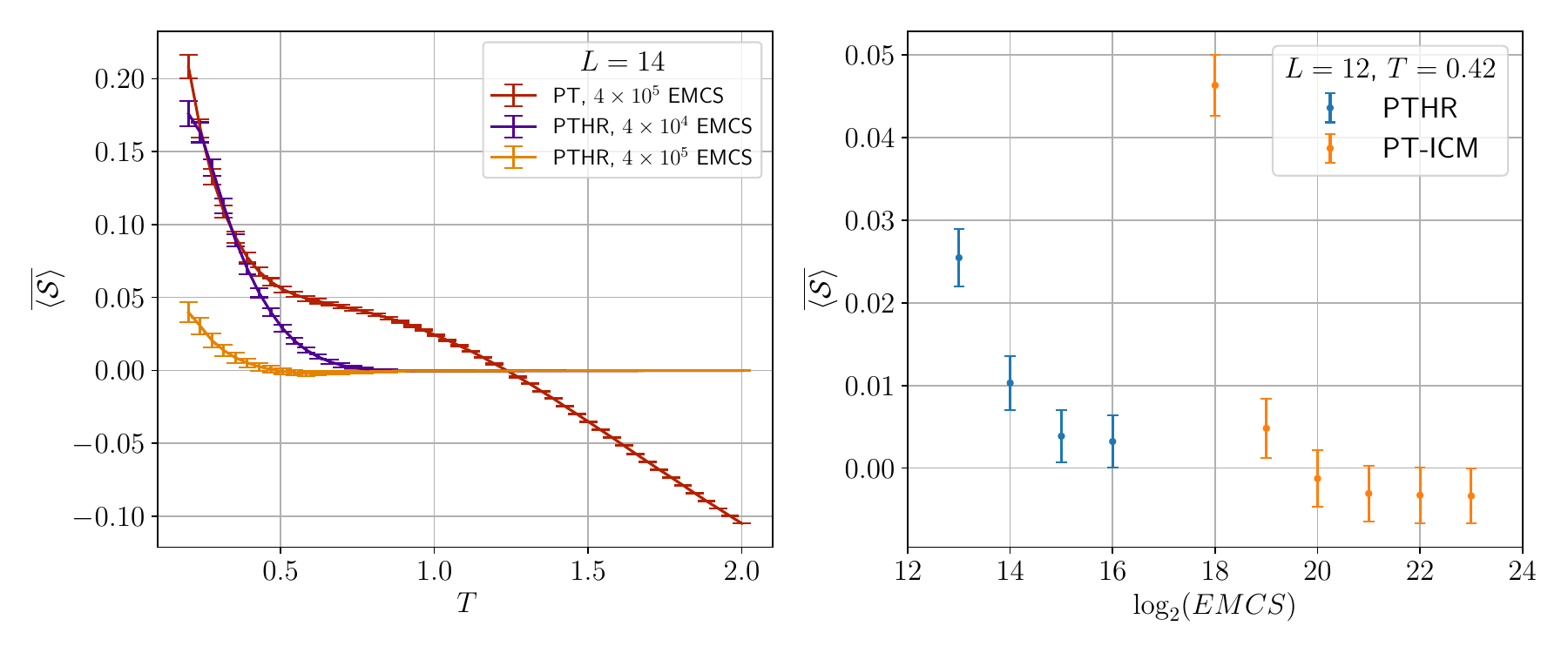}
    \caption{\label{fig:speedup}
        \textbf{The PTHR speedup.} {\bf Left:} $\davg{\SDY}$, defined in Eq.\eqref{eq:SDY-def}, against temperature, as computed for the same set of 2000 samples of $L=14$ lattices in three different simulations, which employ either plain Parallel Tempering (PT) or the algorithm explained in Sec.~\ref{sec:true-algorithm} (PTHR). The orange line corresponds to PTHR with  a total simulation $t_\mathrm{max}=4 \times 10^5$ EMCSs, the purple line is for PTHR with $t_\mathrm{max}=4\times 10^4$ and the red line is PT with $t_\mathrm{max}=4 \times 10^5$. In all three cases, we display the results from bin  0, recall Eq.~\eqref{eq:time-bin-def}. {\bf Right:}   $\davg{\SDY}$ at $T_\mathrm{min}=0.42$ as a function of Monte Carlo time, as computed for a set of 2048 samples of $L=12$ lattices simulated with PTHR (blue points) and for a set of 15000 samples simulated with the algorithm of Ref.~\cite{zhu:15} (orange points labelled PT-ICM, data taken from their Fig.~2-bottom). The time unit is the elementary Monte Carlo step (EMCS) for either algorithm, which coincides with the number of PT sweeps. In the case of PTHR, we use the smallest
        MC time included in every time bin, recall Eq.~\eqref{eq:time-bin-def}.
    }
\end{figure}

\subsection{Comparison between PT and PTHR}

Our comparison between plain PT and PTHR for $L=14$ and $T_\mathrm{min}=0.2$ is carried out in terms of  the SDY observable [Eq.~\eqref{eq:SDY-def}, recall
that the numerical estimate of $\davg{\SDY}$ is compatible with zero if the system has equilibrated]. In the following, it will be important to keep in 
mind that $\davg{\SDY}$ monotonically decreases to zero as $t_\mathrm{max}$ increases. 

The EMCSs for our implementations of PT and PTHR differ only in that the ACU step and the entropic reservoir are missing for plain PT (see Sec.~\ref{sec:true-algorithm} for the description of the ACU step and the entropic reservoir).

After $3.2\times 10^6$ EMCSs of the PTHR algorithm we do not find signatures of missing equilibration given the statistical accuracy granted by the
2000 samples in our simulation; see \ref{sec:appendice_tecnico} and Eq.~\eqref{eq:time-bin-def}.  Although we have been able to equilibrate 746 samples of $L=14$ using plain PT (in the sense that $\davg{\SDY}$ is 
compatible with zero\footnote{We had to discard the first $7\times 10^6$ EMCSs for that. The EMCSs of this plain PT simulation consisted of 10 full-lattice sweeps, followed by a single PT.}), it would be unfair to compare the two algorithms in this way, given the different 
statistical accuracies afforded by the different number of samples. We  have thus followed a different strategy to obtain a more controlled  comparison. 

Specifically, we carry out a PT simulation and identify the value of $t_\mathrm{max}$ for a PTHR simulation that produces equivalent (positive) values of $\davg{\SDY}$. In Fig.~\ref{fig:speedup}--left, we show
$\davg{\SDY}$ as computed for bin~0 in three different simulations. We have a plain PT simulation with $t_\mathrm{max}=4\times 10^5$
and two PTHR simulations, one with $t_\mathrm{max}=4\times 10^5$, the other with $t_\mathrm{max}=4\times 10^4$. All three simulations are conducted on the \emph{same} set of 2000 
samples.  For $T_\mathrm{min}=0.2$, the $\davg{\SDY}$ for $t_\mathrm{max}^\mathrm{PT}=4\times 10^5$ is slightly above the estimate corresponding to 
$t_\mathrm{max}^\mathrm{PTHR}=4\times 10^4$. Hence, the speedup factor seems to be slightly above 10 for $L=14$ and $T=0.2$. However, the speedup factor varies very significantly as a function of temperature. Indeed, while the plain PT simulation seems to be out of equilibrium at all temperatures, including the 
highest  ones, we obtain $\davg{\SDY}$ compatible with zero for  $T\gtrsim 0.8$ in the PTHR simulation. 

One should be careful with these comparisons, however. When one looks in greater detail, PTHR turns out to provide a more significant speedup for the energy than it does for $\ql$ (data for bin~0 and $T=0.2$):
\begin{eqnarray}
t_\mathrm{max}^\mathrm{PT}&=&4\times 10^5 :\qquad       \davg{H}/V=-1.5251(6)\,,\quad  \davg{\ql}=0.8567(16)\,,\\
t_\mathrm{max}^\mathrm{PTHR}&=&4\times 10^4:\qquad      \davg{H}/{V}=-1.6977(3)\,,\quad \davg{\ql}=0.8516(17)\,,\\
t_\mathrm{max}^\mathrm{PTHR}&=&4\times 10^5:\qquad       \davg{H}/{V}=-1.6978(3)\,,\quad  \davg{\ql}=0.8788(14)\,.
\end{eqnarray}

\subsection{Comparison between PTHR and PT-ICM}

The comparison with the PT-ICM algorithm, shown in Fig.~\ref{fig:speedup}--right, is also carried out in terms of the numerical estimate of $\davg{\SDY}$.
Let us stress that we do not carry out new PT-ICM simulations because the relevant results are directly available from Ref.~\cite{zhu:15}. We use as time unit the EMCS for each of the two algorithms (in practice, in units of PT sweeps).

The PTHR datapoints are an average over 2048 samples, as obtained from a simulation with $L=12$, $t_\mathrm{max}=2^{17}$, and $N_T=24$ 
evenly-spaced temperatures between $T_\mathrm{min}=0.42$ and $T_\mathrm{max}=1.8$. We wanted to get the same
conditions of Ref.~\cite{zhu:15}, namely $N_T=26$, but technical constraints in our multispin coding computer program  require  $N_T$ to be a multiple of 4;
see \ref{sec:appendice_tecnico} for further details. 
We opted then for a slightly smaller number of temperatures (mind, however, the 4 additional clones in the entropic reservoir at $T_\mathrm{max}$), which in principle should
put us at a disadvantage due to the slightly reduced acceptance in the PT step.

We find that bin  1 from the PTHR simulation [which includes MC times as
short as $2^{15}$, recall Eq.~\eqref{eq:time-bin-def}] is clearly in equilibrium ($\davg{\SDY}$ is 1 error bar over zero for this set of samples). Instead,
the PT-ICM needs some $2^{21}$ EMCSs to reach their equilibrium value, which is one error bar below zero for their particular set of samples.\footnote{Interestingly enough, $\davg{S}$ from PT-ICM is already compatible with zero, yet \emph{positive}, after only $2^{19}$ EMCS. However, it is clear from Fig.~\ref{fig:speedup}--right that the equilibrium value of $\davg{S}$ for these samples is \emph{negative}.} Hence, the speedup with respect to PT-ICM (as measured in EMCSs for both algorithms) is a factor of $2^6=64$ for $T_\mathrm{min}=0.42$ and $L=12$.

Probably, the reason underlying the large speedup factor of PTHR as compared to PT-ICM may be inferred from Fig.~\ref{fig:cluster-zoology}: $T_\mathrm{min}=0.42$ is near the maximum size for the second-largest clusters (for  $L=14$). As a consequence, the \textit{all}-cluster move in PTHR flips the second-largest cluster
once every two EMCSs (on average). Instead, in a single-cluster algorithm, like PT-ICM, the second cluster would rarely be flipped because it is much smaller than the largest one (either from the majority or from the minority; see  Secs.~\ref{subsec:extended-Intro} and~\ref{sec:true-algorithm} for definitions).

Finally, let us comment on the relative size of the errors in Fig.~\ref{fig:speedup}--right.  The reader will probably be surprised by the similar size of the statistical errors for both algorithms, because the authors of Ref.~\cite{zhu:15} simulated 15000 samples (to be compared with the 2048 samples in the PTHR simulation). The only explanation we 
could find for this discrepancy is that the authors of Ref.~\cite{zhu:15} were computing the disorder average of $\SDY$ at a fixed Monte Carlo time, while 
we are averaging over a time window whose length decreases for higher-numbered bins, recall Eq.~\eqref{eq:time-bin-def}. Indeed, the reader will note that the PTHR errors are identical for 
bins~0 and~1, which means that for those bins the thermal contribution to the error is negligible as compared the sample-to-sample fluctuations (the two types of fluctuations add in quadrature to give the total error). On the other hand, a 
single-time estimate of $\davg{\SDY}$ would have a significant thermal contribution which, we believe, accounts for the similarity of the errors for both 
data sets.

\section{Temperature chaos and the cluster revolution for the temperature random walk}\label{sec:pthr-vs-trw}
\begin{figure}
    \centering
    \includegraphics[width=\linewidth]{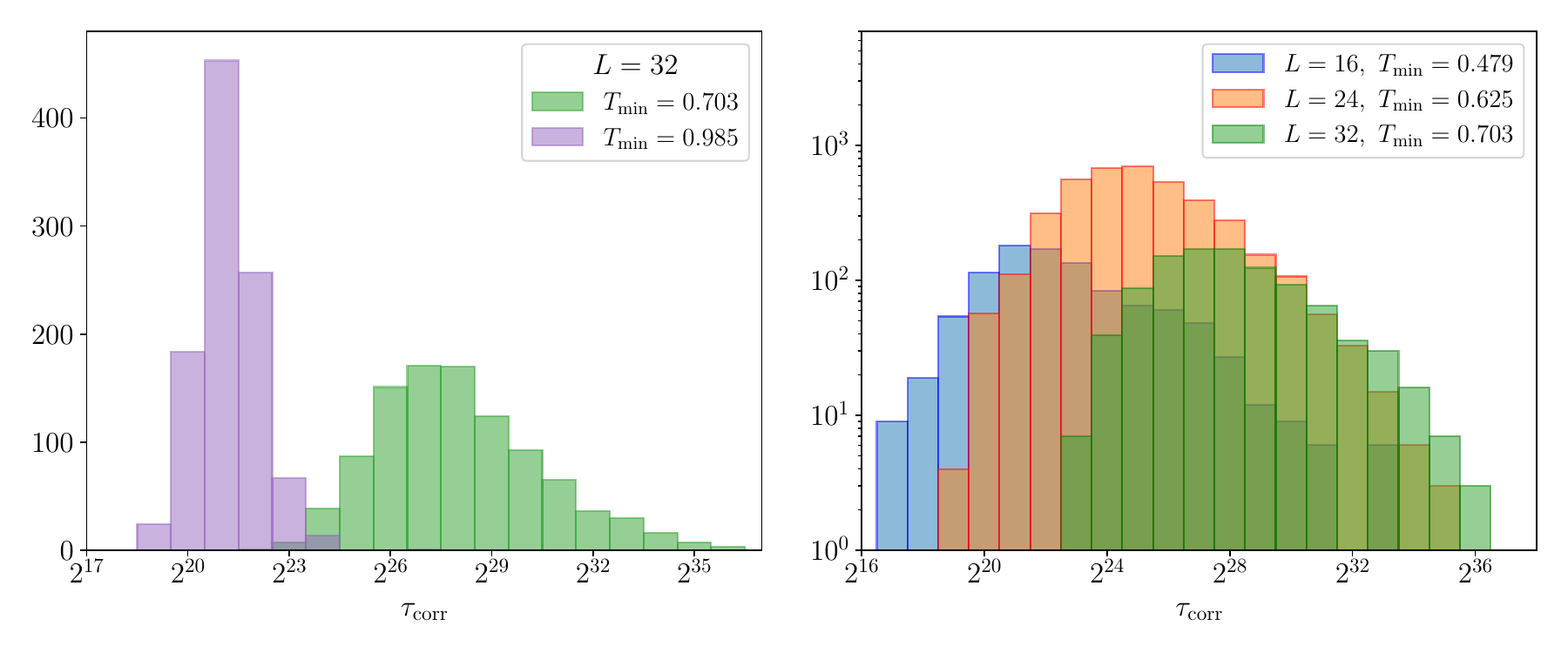}
    \caption{\label{fig:taucorr-old}
        \textbf{Autocorrelation times for plain PT.}  Results of plain PT simulations for different minimum temperature $T_\mathrm{min}$ and different linear sizes $L$ 
        of the 3D Edwards-Anderson model with binary couplings $J_{ij}=\pm 1$. Notice that for binary couplings $T_\mathrm{c} =1.102(3)$~\cite{janus:13}.
        \textbf{Left:} Histogram of the autocorrelation time $\tau_\mathrm{corr}$ for 1000 samples of $L=32$ with $T_\mathrm{min}=0.985\approx0.895T_\mathrm{c}$ (purple) and $T_\mathrm{min}=0.703\approx 0.64 T_\mathrm{c}$ (green).
        \textbf{Right:} Histogram of  autocorrelation times for 1000 samples of $L=32$ with  $T_\mathrm{min}=0.703\approx 0.64 T_\mathrm{c}$ (green), for 4000 samples of $L=24$ with $T_\mathrm{min}=0.625 \approx 0.57T_\mathrm{c}$ (orange) and for 4000 samples of $L=16$ with $T_\mathrm{min}=0.479\approx 0.44 T_\mathrm{c}$ (blue). Data taken from Ref.~\cite{janus:10}.
    }
\end{figure}

Before describing our new PTHR results, let us remind the reader about two major features of the TRW for plain PT simulations. The first feature is that the mixing time for the TRW is \emph{extremely} sample dependent. This property can be quantified by computing the temporal autocorrelation of the temperature index and thence extracting a characteristic time $\tau_\mathrm{corr}$  for each sample~\cite{fernandez:09b,janus:10,yllanes:11}. Figure~\ref{fig:taucorr-old} shows the result for some 
of the plain PT simulations of Ref.~\cite{janus:10}. For instance,
in the 1000-sample simulation for $L=32$ and $T_\mathrm{min}\approx 0.64 T_\mathrm{c}$ one finds samples whose mixing times differ  by a factor 
$\approx 3.2\times 10^4$. The spread of the probability density function of  the mixing times increases as $T_\mathrm{min}$ decreases and/or as 
$L$ increases; see also Fig.~3 in Ref.~\cite{billoire:18}. The second distinctive feature is that this $\tau_\mathrm{corr}$ provides an 
excellent estimate of the time needed to equilibrate a 
given sample using a plain PT algorithm~\cite{fernandez:09b,janus:10,billoire:18}. Furthermore, an unusually large mixing time is  a telltale 
sign   of the presence of  temperature chaos for the sample currently under investigation~\cite{fernandez:13,fernandez:16,billoire:18}. In our 
context, temperature chaos means that there are significant differences between equilibrium spin configurations obtained for temperatures that 
are close in the PT temperature grid. We shall revisit temperature chaos below.

\begin{figure}[t]
    \centering
    \includegraphics[width=\linewidth]{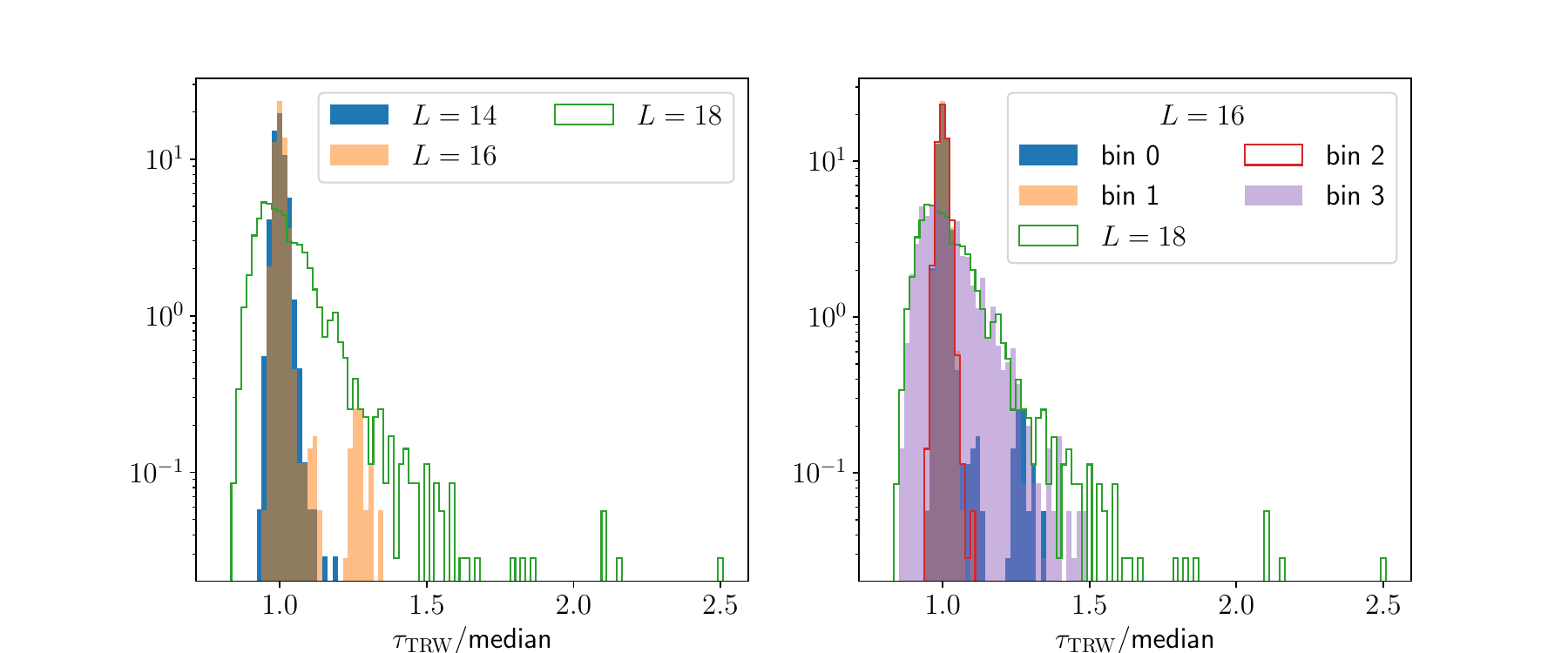}
    \caption{\label{fig:taus_size}
        \textbf{Sample-to-sample fluctuations of the mixing time for the temperature random walk.} Both panels display the probability density 
        function (pdf) for $\tau_\mathrm{TRW}$, see Eq.~\eqref{eq:tautrw}, as computed from the different samples in our simulations, see 
       \ref{sec:appendice_tecnico}. We give $\tau_\mathrm{TRW}$ in units of the median for the corresponding distribution. The 
        different values of the median can be found in Tables~\ref{tab:nrtrps-L16} and~\ref{tab:tautrw-zeff}. We use a logarithmic scale in the 
        vertical axis in order to enhance the visibility of outliers. {\bf Left}: Pdfs for bin~0 of the simulations, recall 
        Eq.~\eqref{eq:time-bin-def}, as computed for lattice sizes $L=14, 16$ and $18$. We know from Eq.~\eqref{eq:SDY} that the simulation for 
        $L=18$ has not
        reached thermal equilibrium. {\bf Right:} Pdfs for the different bins of the $L=16$ simulation and for the bin 0 of the $L=18$ simulation. 
    }
\end{figure}

With the introduction of the ER, which  resets the history of the configurations that get injected into it, the mixing time $\tau_\mathrm{corr}$ extracted from a temporal autocorrelation function is no longer adequate. We introduce instead a new unit of time: the length of the round trip. We define a round trip as the interval of the simulation in which some clone, starting from the $T_\mathrm{max}$ of the PTHR ---\emph{i.e.}, after being injected into the PT chain from the ER---, 
touches the minimum temperature $T_\mathrm{min}$ and returns again to  $T_\mathrm{max}$, where it will be swapped with a reservoir configuration.
We then define the typical timescale for the TRW as the averaged length of the round trip, operatively computed as
\begin{equation}\label{eq:tautrw}
    \tau_\mathrm{TRW} = \frac{N_\mathrm{r} \cdot N_T\cdot \mathrm{number\ of\ EMCSs}}{\mathrm{total\ number\ of\ round\ trips}} \, ,
\end{equation}
where the $N_\mathrm{r}N_T $ factor  accounts for the number of clones trying to achieve a round trip.

The cluster move in PTHR generates a probability density function for $\tau_\mathrm{TRW}$ ---as computed over the different samples--- that barely fluctuates from sample to sample, see Fig.~\ref{fig:taus_size}--left. Rather than the orders-of-magnitude fluctuations 
characteristic of plain PT~\cite{janus:10,fernandez:16,billoire:18}, we observe for PTHR that the $\tau_\mathrm{TRW}$ of any of our samples differs from the median of the distribution by 
less than a factor of three. As a consequence, for the PTHR algorithm, we can meaningfully identify a characteristic time scale for the TRW, namely the median of the distribution, that applies to \emph{all} samples. Another nice feature of the probability density 
function for $\tau_\mathrm{TRW}$ is that its form can be inferred also from not fully equilibrated data. Indeed, see 
Fig.~\ref{fig:taus_size}--right, for $L=16$ only the distribution for bin  3 is clearly wider than its equilibrium counterpart.

So far, so good, but when one compares the median $\tau_\mathrm{TRW}$ in Table~\ref{tab:tautrw-zeff} with the number of EMCSs needed to reach thermal 
equilibrium, a mismatch of three orders of magnitude between timescales is self-evident; see \ref{sec:appendice_tecnico}. In other words, and 
at variance with plain PT, the mixing time for the TRW does not identify the equilibration time for PTHR.  

\subsection{On the hard-to-equilibrate samples and temperature chaos}\label{subsect:bad-samples}

The results of the previous section make one wonder whether a small subset of samples is responsible for the  slow equilibration, as is the case in PT simulations.

\begin{figure}[t]
    \centering
    \includegraphics[width=\linewidth]{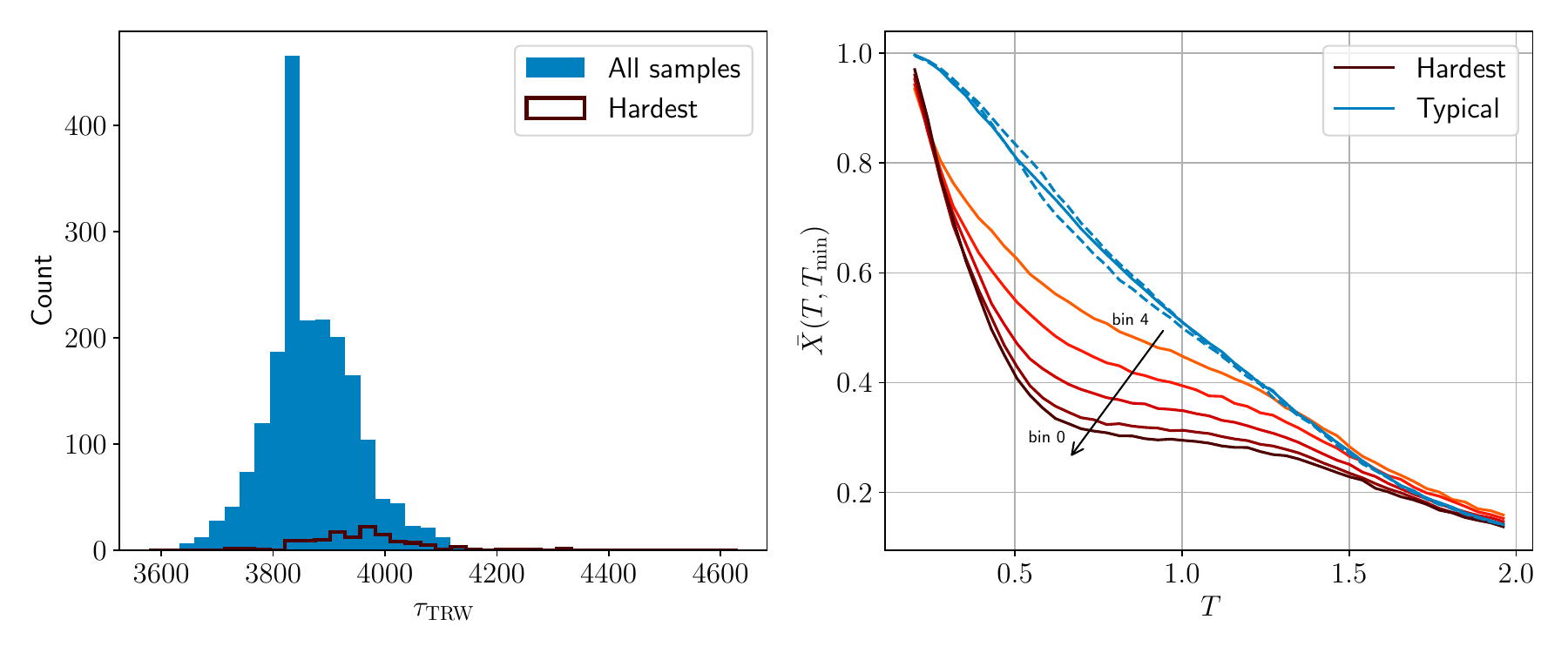}
    \caption{\label{fig:trw-chaos}
        \textbf{Temperature random walk and temperature chaos.} This figure shows the difference between typical samples and the hardest samples we found in  our  simulations 
        of 2000 samples of $L=14$ lattices, with  $T_\mathrm{min}=0.2$. The hardest samples were identified following the procedure explained in Sec.~\ref{subsect:bad-samples}.
        \textbf{Left:} Histogram of $\tau_\mathrm{TRW}$ as defined in Eq.~\eqref{eq:tautrw}. The blue histogram is the computed value for the 2000 samples of $L=14$, from 
        which we computed the observables for this size in this work. The superimposed brown histogram is for the 129 hardest samples of the set.
        \textbf{Right}: Chaotic parameter as a function of the temperature  $T$ [the parameter is defined in Eq.~\eqref{eq:caotic}, we fix $T_1=T_\mathrm{min}=0.2$ and let $T_2=T$ vary], averaged over different subsets of 129 samples. For the set of hardest samples, 
        the chaos signal increases as equilibration proceeds from bin  4 to  bin  0, recall Eq.~\eqref{eq:time-bin-def}. As an example of typical behaviour, the blue line is the chaotic parameter averaged over
        129 samples randomly drawn from the set of 2000. More specifically, we performed 50 random extractions of sets of 129 samples and ordered them according to the value of the chaos parameter at $T=0.63$. The curve shown corresponds to the median group of 129 samples. In order to quantify fluctuations, we also draw with dashed blue lines the curves for percentiles 16 and 84.
    }
\end{figure}

We investigated the problem by comparing the thermal averages from two simulations of the same set of 2000 samples of $L=14$ lattices, with $T_\mathrm{min}=0.2$.
In the first 
simulation, which used a preliminary version of the algorithm, we employed only Metropolis, PT and the ACU (the simulation was nevertheless long enough to equilibrate). The 
second simulation, with the standard 
PTHR algorithm, is the one quoted in \ref{sec:appendice_tecnico} with $t_\mathrm{max}=6.4\times 10^6$. Then, we computed for every sample 
the difference between the thermal averages of $e\equiv H/V, \ql$ and $q^2$, as obtained for bin  3 of the PTHR simulation, with their partner from the first simulation.  Let  $\Delta e(s)$, $\Delta \ql(s)$ and $\Delta q^2(s)$ be the difference between our two estimates of thermal averages for sample $s$. Should
thermal equilibrium have been reached for bin 3, the sample averages over the $N_\mathrm{s}=2000$ samples
\begin{equation}
\hspace{-2cm}\overline{\Delta e(s)}=\frac{1}{N_\mathrm{s}}\sum_{s=1}^{N_\mathrm{s}}\, \Delta e(s)\,,\quad
\overline{\Delta q_l(s)}=\frac{1}{N_\mathrm{s}}\sum_{s=1}^{N_\mathrm{s}}\, \Delta q_l(s)\,,\quad
\overline{\Delta q^2(s)}=\frac{1}{N_\mathrm{s}}\sum_{s=1}^{N_\mathrm{s}}\, \Delta q^2(s)\,,
\end{equation}
would have been compatible 
with zero. Note that it is straightforward to compute errors for the above three disordered averages, which are very accurate because the contribution to the error of the sample-to-sample fluctuations is suppressed to a large degree by our use of the \emph{same} set of 2000 samples in the two simulations.
Of course, since bin  3 of the PTHR simulation is not yet thermalized, we found $\overline{\Delta e(s)}>0$ , $\overline{\Delta \ql(s)}<0$, and $\overline{\Delta q^2(s)}<0$. Although $\overline{\Delta e(s)}$  is the smallest difference in absolute value (as expected), it becomes the largest when all three differences are divided by their respective
statistical errors. Hence, we focused our attention on the energy difference $\Delta e(s)$ as the most sensitive quantity to lack of equilibration. Next, we reordered the sample labels in such a way that 
\begin{equation}
    \Delta e(s_1)<\Delta e(s_2)< \Delta e(s_3)<\ldots< \Delta e(s_{N_\mathrm{s}})\,.
\end{equation}
The lack of equilibration implies that 
$\sum_{s=1}^{N_\mathrm{s}}\Delta e(s)=\sum_{i=1}^{N_\mathrm{s}} \Delta e(s_i) > 0$,
which suggested identifying as  the \emph{non-equilibrated set} all samples $s_k$ with $k\geq i^*$, where $i^*$ is the smallest integer such that $\sum_{i=1}^{i^*} \Delta e(s_i) > 0$. In this way, we 
obtained a set of only 129 non-equilibrated samples.  When we combined in the sample average  the results from bin  0 for these 129 samples and the thermal expectations from
bin  3 for the remaining $N_\mathrm{s}-129$ samples, the equilibration test in Eq.~\eqref{eq:SDY} was passed. This consistency check convinced us that this identification of 
the subset of the hardest-to-equilibrate samples is sensible.

The next question regards the search for some  feature of the hard-to-equilibrate samples that could allow us to identify 
them at the beginning of the simulation. If one succeeds, the potential computational gain is very significant because the simulation of a large fraction of 
the samples could be stopped well before the equilibration condition in Eq.~\eqref{eq:SDY} is met,  without compromising the final results. Our answer  to the question is disappointing: The identifying feature exists (it is temperature chaos), but assessing the strength of temperature chaos for a given sample requires bringing the simulation to equilibrium, which  eliminates the potential computational gain.

A first negative result is in Fig.~\ref{fig:trw-chaos}--left, where we show that the mixing time $\tau_{\mathrm{TRW}}$ would not do a good job of identifying the hard samples. Indeed, the probability density function of the hard samples is slightly displaced to large values of $\tau_{\mathrm{TRW}}$ as compared to the distribution for the whole set of samples. Unfortunately, however, some of the hardest samples have a lower-than-average $\tau_{\mathrm{TRW}}$ for the PTHR algorithm.

Then, inspired by our previous experience with plain PT, we turned our attention to temperature chaos~\cite{fernandez:13,fernandez:16,billoire:18}. 
The observable that quantifies temperature chaos in a sample is the \textit{chaotic parameter}:
\begin{equation}\label{eq:caotic}
    X_{T_1, T_2} =  \frac{\tavg{q^2_{T_1, T_2}}}{\sqrt{\tavg{q^2_{T_1, T_1}}\tavg{q^2_{T_2,T_2}}}} \, ,
\end{equation}
where $q^2_{T_1, T_2}$ is the squared overlap [Eq.~\eqref{eq:tot_ovlp}] computed between clones at different temperatures $T=T_1, T_2$. We regard the chaotic parameter as a correlation coefficient:
if $X_{T_1,T_2}$ is near one, then the spin configurations at $T_1$ and $T_2$ are structurally similar. The smaller $X_{T_1,T_2}$, the more structurally different the configurations that are typical at both temperatures.

Fig.~\ref{fig:trw-chaos}--right shows the behaviour of $X_{T_\mathrm{min}, T}$ as a function of the larger temperature $T_2=T$, averaged over the  129 
hardest samples from our data set (the averaging significantly smooths the curve). As expected, the curve for the hard samples lies significantly below the curve for 129 samples extracted with uniform probability (blue line). Thus, temperature chaos is stronger for the hardest samples than it is for the average sample. We infer that, 
despite the dramatically enhanced mobility of the clones in the PT chain, temperature chaos is still  the limiting factor that precludes fast  equilibration for PTHR. Unfortunately, see Fig.~\ref{fig:trw-chaos}--right,  the chaotic signal is much weaker for bin  4 than it is for bin  0, which suggests that strong temperature chaos is not easy to identify from still non-equilibrated data.

\subsection{The round trip as a valid unit of time (after taking the disorder average)}\label{subsec:rtrips}
The mixing time for TRW is clearly much shorter than the equilibration time, but one could wonder whether or not it also scales with system size. Having this question in mind, we recall that our equilibration criterion in Eq.~\eqref{eq:SDY} refers to the whole set of samples in the simulation. Then, we prefer to rephrase our question in a slightly different way: \emph{how many round trips $T_\mathrm{max}\to T_\mathrm{min}\to T_\mathrm{max}$ need to be accomplished by the individual clone during the PT part of the PTHR simulation to ensure that the criterion in Eq.~\eqref{eq:SDY} is met?}
\begin{figure}[t]
  \centering
  \includegraphics[width=0.5\linewidth]{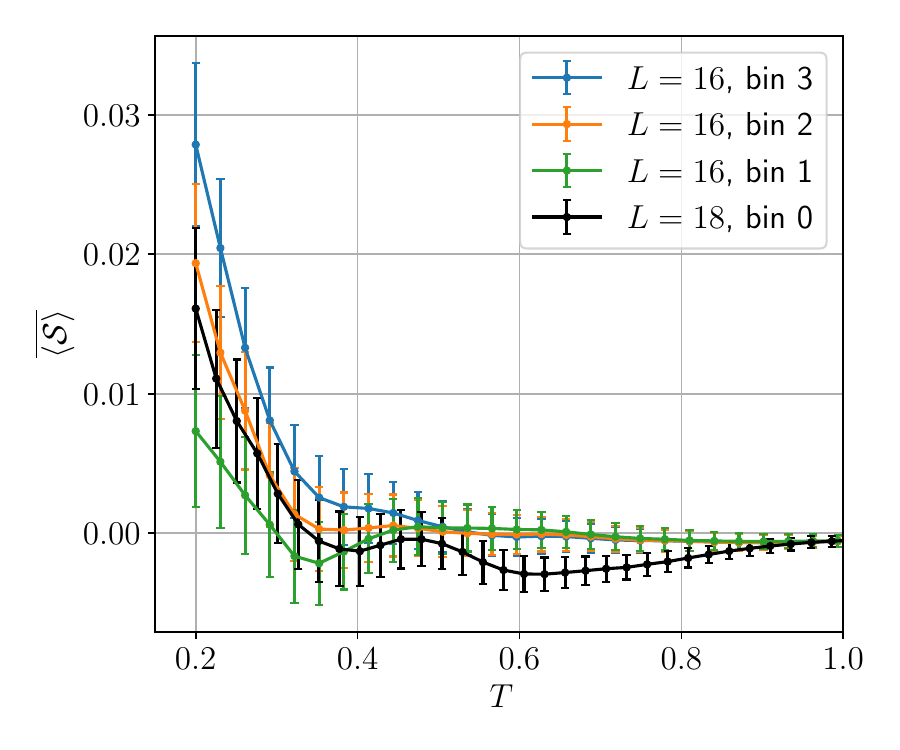}  
\caption{\label{fig:compare_rndtrps-SDY}
        Equilibration Schwinger-Dyson-Young parameter $\davg{\SDY}$ [Eq.~\eqref{eq:SDY-def}] versus temperature, as computed for time bins 1 (green), 2 (orange) and 3 (blue) of our PTHR simulation for $L=16$, $T_\mathrm{min}=0.2$ and $t_\mathrm{max}=3.84\times 10^7$. Time bins are defined in Eq.~\eqref{eq:time-bin-def}. The black curve is obtained from bin  0 of our $L=18$ simulation (see \ref{sec:appendice_tecnico}), which has not achieved equilibration.
}
\end{figure}

\begin{table}[tb]
    \caption{\label{tab:nrtrps-L16}
        Median number of round trips for the  different time bins, recall Eq.~\eqref{eq:time-bin-def},  corresponding to  our PTHR simulation for $L=16$, $T_\mathrm{min}=0.2$ and $t_\mathrm{max}=3.84\times 10^7$. The values shown are the median (over the samples) of the average number of round trips $T_\mathrm{max}\to T_\mathrm{min}\to T_\mathrm{max}$ accomplished by a single clone during the PT part of the algorithm. The mixing time
        $\tau_\mathrm{TRW}$ has been computed sample by sample from Eq.~\eqref{eq:tautrw}. The value shown corresponds to the median $\tau_\mathrm{TRW}$ over the samples. Errors were computed using a bootstrap method.
    }
\centering
\begin{tabular}{c . (}
\hline
\multicolumn{1}{c}{Bin}                                                   &   \multicolumn{1}{c}{Round trips (median)} & \multicolumn{1}{c}{$\tau_\mathrm{TRW}$ (median)} \\ \hline
 $3$    & 271.0(5)                  & 8854(15)          \\
 $2$    & 828.3(4)                  & 5795(3)        \\
 $1$   & 1573.1(6)                 & 6103(2)        \\
 $0$  & 3144.0(14)               & 6105(3)        \\ 
\hline
\end{tabular}
\end{table}

What makes it attractive to consider the number of PTHR round trips ---or, equivalently, $\tau_\mathrm{TRW}$; recall Eq.~\eqref{eq:tautrw}--- as the effective simulation length is that its sample-to-sample fluctuations are negligible by spin-glass standards. The median number of round trips for our $L=16$ simulation is given in Table~\ref{tab:nrtrps-L16}. We see that the typical clone performs $\approx 3150$ round 
trips during bin  0. Since bin 0 corresponds to the second half of the simulation, this typical clone performs  approximately another $3150$ round trips \emph{before} 
bin  0 starts. Hence, all samples in our simulation are sufficiently equilibrated after $\approx 3150$ round trips to pass the equilibration test in 
Eq.~\eqref{eq:SDY}, at least within the accuracy afforded by the approximately 2000 samples in our simulations. One could speculate that the same number of round trips would also suffice to equilibrate other system sizes at $T_\mathrm{min}=0.2$.

In order to assess whether or not the above speculation is sensible, we turn to our non-equilibrated PTHR simulation of the $L=18$ lattice. We estimate that the  typical clone has  performed some 2550 round trips during the whole simulation, or $\approx 1275$ round trips before  data acquisition starts in bin~0. From  this point  of view, see Table~\ref{tab:nrtrps-L16}, bin  0 of the $L=18$ simulation is  intermediate between bins  1 and  2 of the 
$L=16$  simulation. Interestingly enough, also for $\davg{\SDY}$, see Fig.~\ref{fig:compare_rndtrps-SDY}, bin  0 of the $L=18$ simulation is 
intermediate  between bins  1 and  2 of the $L=16$ simulation. The simplest interpretation of this finding is that the PTHR mixing time 
$\tau_\mathrm{TRW}$  does scale with $L$ as the true equilibration time, $\tau$, corresponding to the hardest samples in our data set.

One can use the above intuition to define an effective dynamic exponent for the PTHR algorithm 
\begin{equation}
z^{\mathrm{PTHR}}=\frac{\mathrm{d}\log \tau}{\mathrm{d} \log L}\approx 
\frac{\mathrm{d}\log \tau_\mathrm{TRW}}{\mathrm{d} \log L}\,.
\end{equation}
Should $z^{\mathrm{PTHR}}$ turn out to be independent from $L$, the computational complexity would grow as power law,
$\tau\propto L^{z^{\mathrm{PTHR}}}$. In plain PT simulations, however, the analogous dynamic exponent $z^\mathrm{PT}$ has turned out to grow when 
$L$ increases or when the minimal temperature of the PT chain $T_\mathrm{min}$ decreases.  Hence, we need an effective, $L$-dependent exponent that we obtain from lattices $L_1$ and $L_2$ following Ref.~\cite{billoire:18}:
\begin{equation}\label{eq:z-eff}
 \hspace{-1cm}   z_\mathrm{eff}^\mathrm{PTHR}(T_\mathrm{min}, L_1, L_2) = \frac{\log \Big[\mathrm{median}[\tau_\mathrm{TRW}(L_1)]\Big]-\log \Big[\mathrm{median}[\tau_\mathrm{TRW}(L_2)]\Big]}{\log L_1- \log L_2} \,.
\end{equation}

The results for the effective exponent, along with our estimates of the median $\tau_\mathrm{TRW}$ for $L=14, 16, 18$ are reported in Table~\ref{tab:tautrw-zeff}. 
As a standard of comparison, let us recall that for plain PT one finds  $z_\mathrm{eff}^\mathrm{PT}(T_\mathrm{min}\approx 0.64 T_\mathrm{c};L_1,L_2)>11$
for all $12\leq L_1<L_2$ when one considers the hardest samples (percentile 90 of hardness in simulations with binary couplings; see Fig.~4 of Ref.~\cite{billoire:18}). Therefore, it is remarkable that all values of $z_\mathrm{eff}^\mathrm{PTHR}(T_\mathrm{min}, L_1, L_2)$ in Table~\ref{tab:tautrw-zeff} are
smaller than 7.7, in spite of our much smaller $T_\mathrm{min}\approx 0.21 T_\mathrm{c}$. 

\begin{table}[t]
    \caption{\label{tab:tautrw-zeff}
        Median over samples of $\tau_\mathrm{TRW}$, the mixing time of the temperature random walk of the PTHR algorithm,  for system sizes  $L=14, 16, 18$.  We compute $\tau_\mathrm{TRW}$ from  Eq.~\eqref{eq:tautrw}. Errors were computed using a bootstrap method. 
        The effective exponent $z_\mathrm{eff}$ defined in Eq.~\eqref{eq:z-eff} quantifies the growth of the equilibration time with the system size.  }
\centering
\begin{tabular}{ c ( . }
\hline
\multicolumn{1}{c}{$L$}  & \multicolumn{1}{c}{$\tau_\mathrm{TRW}$} & \multicolumn{1}{c}{$z_\mathrm{eff}^\mathrm{PTHR}(L_1=L-2, L_2=L)$ }\\ \hline
$14$ & 3871(1)           &      \multicolumn{1}{c}{---} \\
$16$ & 6105(3)           & 3.412(2)         \\
$18$ & 15\,050(40)         & 7.6588(15)          \\ 
\hline
\end{tabular}
\end{table}

\section{Conclusions}\label{sect:conclusions}

We have obtained an effective cluster method for the simulation of three-dimensional spin glasses down to very low temperatures by introducing some simple modifications of  Houdayer's cluster 
algorithm~\cite{houdayer:01}. 
Under some  conditions, our algorithm outperforms previous proposals~\cite{zhu:15} by a speedup factor $\approx 64$.\footnote{Let us stress that we are talking about cubic lattices. For very elongated prisms, the performance of the cluster method is far better than it is in cubic lattices~\cite{bernaschi:26b}.} We have related this large speedup to the sizes
of the second and third clusters ---as ordered by their size--- for both signs of the overlap. These clusters are flipped with much larger frequency using the all-cluster scheme advocated here than they are 
with single-cluster update schemes~\cite{houdayer:01,zhu:15}. An unexpected feature of the algorithm is that the temperature random walk performed during the PT part of the algorithm 
is \emph{extremely} fast compared to plain PT simulations~\cite{fernandez:09b,janus:10, billoire:18}. In particular, the orders-of-magnitude sample-to-sample fluctuations for the mixing time of the temperature random walk completely disappear.

We should mention two limitations. First, the sizes of the second and third clusters have a maximum at temperatures $T\approx 0.4 T_\mathrm{c}$. This is the temperature where the maximum 
speedup is expected. Unfortunately, the sizes of the clusters diminish for smaller temperatures. Second, although we have ascertained that equilibration time varies significantly from sample to sample, 
we have not found any way to identify early in the simulation those samples that are harder to equilibrate. Temperature chaos has turned out to be a distinctive feature of hard samples, but this property 
is difficult to asses unless thermal equilibrium is reached, which makes it impractical for the early identification of hard samples.
Instead, our equilibration criterion relies on a disorder-averaged Schwinger-Dyson-Young identity. Therefore, the whole set of samples have to be equilibrated before the hard instances can be identified, meaning that most samples will be simulated for far longer times than needed and preventing us from concentrating computational effort on the few hard cases.

Overall, the new algorithm has allowed us to break a world record in the size of equilibrated systems down to $T_\mathrm{min}=0.2\approx 0.21 T_\mathrm{c}$, which has turned out to be quite 
important for our physics investigation~\cite{chilin:26}.

\ack{We warmly thank Prof. Luis Antonio Fernández for many useful discussions and insights. We also thank Prof. Federico Ricci-Tersenghi and Dr. Luca Maria del Bono for discussions. We are also thankful to Prof. Jon Machta for useful correspondence.

This work was partially supported by Ministerio de Ciencia, Innovación
y Universidades (Spain) and by the European Regional Development Fund
through grants no.
PID2022-136374NB-C21, PID2024-156352NB-I00, PID2024-158623NB-C22, and
RED2022-134244-T (MCIU/AEI/10.13039/501100011033/FEDER, UE); by Junta de Extremadura (Spain) through grant no.~GR24022;
and by funding from the 2021 first FIS (Fondo Italiano per la Scienza) funding
scheme (FIS783 - SMaC - Statistical Mechanics and Complexity) from Italian MUR
(Ministry of University and Research).
 We acknowledge the use of the CESAR computational resources at the BIFI Institute (University of Zaragoza) and those of the Instituto de Computacion Cient\'{\i}fica Avanzada de Extremadura (ICCAEx).}

\section*{Data availability}

The data contained in the figures in this paper, as well as the Jupyter notebook that generates them, can be downloaded from \href{https://github.com/JeanShc/pthr}{https://github.com/JeanShc/pthr}.

\appendix
\setcounter{figure}{0}
\section{Technical details on the simulations}\label{sec:appendice_tecnico}
\counterwithin{figure}{section}
To make the most of our computational budget,
we were concerned not only with the algorithmic scaling with the system size $L$ and the number of systems $N_\mathrm{r}N_T$, but also with reducing the total CPU time needed for the simulations. We have mostly attempted to accelerate step (i) of our algorithm, namely the full-lattice update ---see the description of Sec. \ref{sec:true-algorithm}---, which is the costliest one in CPU time. Our chosen approach is multispin coding, assisted with look-up tables (LUTs); see \ref{subsec:MSC}.
The parameters of our simulations are given in~\ref{subsec:parameters}.

\subsection{Multispin coding for Gaussian couplings through look-up tables}\label{subsec:MSC}

Multispin coding~\cite{jacobs:81} is a very efficient technique to simulate several system in parallel, by profiting from the fact that boolean operations ({\tt AND}, {\tt XOR}, {\tt OR}, etc.) are carried out simultaneously and independently for all bits in a computer word. If one notices that our spins $\sigma_i=\pm 1$ can take only two values, it comes naturally to identify a spin with a single bit within the computer word [\emph{i.e.}, a binary variable $b_j=(1+\sigma_i)/2=0,1$]. However, multispin coding is most efficient when the couplings in the Hamiltonian are $J=\pm 1$, because in that case the energy increment caused by a spin flip can take only a small number 
of  values (for a hypercubic lattice in $D$ dimensions $\Delta E=0,\pm 4,\ldots\pm 4D$).  In this simple setting it is relatively straightforward to find a few boolean operations 
that implement exactly the (say) Metropolis algorithm; see for instance Ref.~\cite{newman:99}. The problem becomes more complicated when the energy change $\Delta E$ is a real variable, as it is the case for Gaussian couplings.

The above difficulty has been efficiently solved in the past~\cite{fernandez:16b} using a LUT. Unfortunately, the solution of Ref.~\cite{fernandez:16b} is quite specific to the 
 square lattice's connectivity~4: the neighbourhood of the spin that is being updated is determined by just four binary couplings, hence $2^4=16$ different possibilities per lattice site. This results 
in short LUTs and in a small number of boolean operations that decide whether or not the spin should be flipped. In $D=3$, the local neighbourhood of a site consists of 6 spins and hence $2^6=64$ different 
possibilities per site. Furthermore, the necessary time to equilibrate may vary  significantly from sample to sample, recall Section~\ref{subsect:bad-samples}. Thus, we have preferred to avoid multispin-coding solutions that 
implement different samples in a single computer word.

Our starting observation is that the energy change upon flipping spin $j$ is  given by
\begin{equation}
    \Delta E = 2 \sum_{j\in \partial_i} \sigma_i J_{ij}\sigma_j\,
\end{equation}
where $\partial_i$ is the set of the 6 lattice nearest-neighbours of site $i$. Hence, $\Delta E$ is determined by the six binary variables $\sigma_i\sigma_j$ that can be coded in a single byte. Given
that a computer word is typically composed of 8 bytes, we chose to code 8 different systems in a computer word (namely, 4 clones from two different replicas, see Sec.~\ref{sec:true-algorithm} for terminology). All 8 systems in a word correspond to the same lattice site $i$, and hence share the value of the six couplings $J_{ij}$.

In its most ambitious form, our program stores in a LUT the 64 possible values of the Metropolis acceptance probability $\min\{1,\exp(-\Delta E/T)\}$. The 8 systems in a computer word compare their
acceptance probability with 8 independent random numbers. Specifically we use streaming extensions and 256-bit computer words to obtain a parallel implementation of  8 independent {\tt xoshiro256++} pseudorandom number generators~\cite{blackman:21}. This approach is extremely effective for small systems. However, it  requires storing $64 N_T$ different probabilities per site in the LUT  (recall that 
$N_T$ is the number of temperatures in the PT grid). The LUT size quickly grows with $L$ because $N_T\propto L^{D/2}$. Furthermore, the 8 systems coded in the same 
computer word are at a different temperature, hence the $64 N_T$ probabilities in the LUT need to be accessed in a random order. It is then unsurprising that the efficiency of this approach deteriorates 
 with system size. 

The above difficulty explains  why  we have opted for a much smaller energy LUT for our largest systems. In the energy-LUT approach we store the 64 possible values of $\Delta E$ for lattice site $i$.
In the case of Metropolis, the value of $\Delta E^{(k)}$ is divided by $T^{(k)}$ and compared with the logarithm of the corresponding random number $R^{(k)}$ (the superscript $k$ labels each of the 8 systems coded 
in the computer word corresponding to site $i$). We have found that the overhead of computing the logarithm is compensated by the faster memory access to a smaller LUT. In the case of Microcanonical 
Simulated Annealing (MicSA)~\cite{bernaschi:26}, it is even simpler because we just need to compare $\Delta E^{(k)}$ with the energy stored in the walker. If the walker has enough energy, the spin is flipped 
and the energy $\Delta E^{(k)}$ is subtracted from that system's walker.

Nevertheless, we have in all cases used the probability LUT for the clones in the ER, since all 8 systems in the reservoir are at the same temperature. Indeed, the reader will have noticed that the entire ER can be used with a single computer word per lattice site.

\subsection{Parameters used in our simulations}\label{subsec:parameters}

Our local update rule in step  (i) of our algorithm, recall Sec.~\ref{sec:true-algorithm}, 
has been Metropolis for  the simulations at $L=12, 14$ and for bins 2 and 3 of $L=16$. For bins 0 and 1 of $L=16$ and for the $L=18$  simulations we used MicSA~\cite{bernaschi:26} instead [recall definition of bin in Eq.~\eqref{eq:time-bin-def}].

Table \ref{tab:PTparameters} reports the simulation parameters. 
The PT uses $N_T$ temperatures uniformly distributed between $T_\mathrm{min}$ and $T_\mathrm{max}$ (both included).

\begin{table}[h]
\caption{ \label{tab:PTparameters}
    Parameters used for the simulations. 
    PTHR = Parallel Tempering with Houdayer moves and entropic  Reservoir, using Metropolis for local updates, PTHR* = Parallel Tempering with Houdayer moves and entropic Reservoir, using MicSA for local updates.  In the $L=16$ case, 
    the simulation used PTHR for bins  2 and  3 and then switched to PTHR*.}
\centering
\begin{tabular}{lllllll}\hline
$L$ \ \ & $T_\mathrm{min}$ \ \ & $T_\mathrm{max}$  \ \ & $N_T$ \ \ &   samples \ \ & $t_\mathrm{max}$         \ \ & Algorithm  \\ \hline
$12$  & $0.42$      & $1.8$       & $24$        & $2048$       & $2^{17}$             & PTHR       \\
$14$  & $0.2$       & $2$         & $48$        & $2000$       & $6.4 \times 10^6$  & PTHR       \\
$16$  & $0.2$       & $2$         & $60$        & $2081$       & $3.84 \times 10^7$ & PTHR/PTHR* \\
$18$  & $0.2$       & $2$         & $72$        & $2028$       & $3.48 \times 10^7$ & PTHR*      \\
 \hline
\end{tabular}
\end{table}
\section*{References}

\providecommand{\newblock}{}

\end{document}